PRELIMINARY CHARACTERIZATION OF AN ACTIVE SPIRAL GALAXY NEAR M57

Author: Elio Quiroga Rodríguez
Universidad del Atlántico medio (UNAM), Faculty of Communication
Universidad Internacional de Valencia (VIU)
**elio.quiroga@pdi.atlanticomedio.es**

ABSTRACT

M57 es una de las nebulosas planetarias más observadas y populares. Recientemente, el Telescopio James Webb la ha fotografiado, obteniendo unas imágenes fascinantes (NASA, 2023). Como pasa con otros objetos muy conocidos y populares, en sus cercanías hay otros objetos astronómicos de interés que en ocasiones pueden pasar desapercibidos. Muy cerca de esta nebulosa (en las coordenadas 18:53:18.81 +33:03:59.89 -en j2000-) hay una interesante galaxia espiral de tipo SBbc, conocida como IC 1296. A más distancia angular de M57 e IC 1296, en las coordenadas 18:53:09.625 +33:05:38.85 hay un objeto muy tenue, llamado WISEA J185309.64+330538.7. El autor presenta los resultados preliminares de los análisis de los espectros de ambas galaxias, obtenidos en el telescopio griego de Skinakas, presentando algunas conclusiones sobre su tipo, así como distancia en el caso de WISEA J185309.64+330538.7. Las dos galaxias muestran trazas de tener núcleos galácticos activos, y WISEA J185309.64+330538.7 es una radiogalaxia con características Seyfert 1, mientras que IC 1296 muestra algunas líneas que podrían indicar su pertenencia a la categoría intermedia Seyfert 1.5 (entre Seyfert 1 y 2), además de ser un conocido emisor de rayos X.

ABSTRACT

M57 is one of the most observed and popular planetary nebulae. Recently, the James Webb Space Telescope photographed it, capturing fascinating images (NASA, 2023). As is the case with other well-known and popular objects, there are other nearby astronomical objects of interest that can sometimes go unnoticed. Very close to this nebula (at coordinates 18:53:18.81 +33:03:59.89 -in j2000-), there is an intriguing spiral galaxy of type SBbc, known as IC 1296. At a greater angular distance from M57 and IC 1296, at coordinates 18:53:09.625 +33:05:38.85, there is a very faint object called WISEA J185309.64+330538.7. The author presents preliminary results of the spectral analyses of both galaxies, obtained at the Skinakas Observatory (Greece), providing some conclusions about their types and, in the case of WISEA J185309.64+330538.7, its distance. Both galaxies exhibit traces of having active galactic nuclei, with WISEA J185309.64+330538.7 being a radio galaxy with Seyfert 1 characteristics, while IC 1296 shows some lines that could indicate its membership in the intermediate category Seyfert 1.5 (between Seyfert 1 and 2), in addition to being a known X-ray emitter.

DISCUSSION

In the images from the HST Proposal 12309, near M57, a relatively well-known galaxy, IC 1296, a spiral of type SBbc (Steinicke & Jakiel, 2007), can be observed at coordinates 18:53:18.81 +33:03:59.89. It is classified as a Low Surface Brightness Galaxy in Simbad[1], with a magnitude of 14.80 +/- 0.30 in NED[2]. It is a coveted target for amateur astronomers due to its low brightness. Its distance (approximately 73.58 Mpc) and redshift (0.01708 +/- 0.00003), as well as other characteristics, are available in public catalogs. On August 18, 2013, a supernova was observed in IC 1296, known as SN2013ev (Ciabattari et al., 2013).

---

1 http://simbad.cds.unistra.fr/simbad/sim-basic?Ident=IC+1296&submit=SIMBAD+search
2 http://ned.ipac.caltech.edu/byname?
 objname=IC+1296&hconst=67.8&omegam=0.308&omegav=0.692&wmap=4&corr_z=1

At the coordinates 18:53:09.625 +33:05:38.85, the object known as WISEA J185309.64+330538.7 or 2MASX J18530959+3305385 is listed in the Wide-field Infrared Survey Explorer (WISE) catalog (Anderson et al., 2014) and the Two Micron All-Sky Survey Extended Source Catalog of 2MASX (Karachentseva et al., 2010). In both catalogs, it is categorized as an Infrared Source (IrS), as well as in SSTSL2[3] and 2MASS (Cutri et al., 2003). It is listed as a galaxy in the LEDA[4] and CGMW catalogs (Saito et al., 1990). In an image captured by the Hubble Space Telescope (HST Proposal 12309, O'Dell, 2010) with the WFC3 camera, WFC.F658N filter (Hα), it is located at the edge of the image, revealing an intriguing albeit faint barred spiral shape, which the author identifies as type SBc based on the HST image, as shown in Figure 1f. To the best of the author's knowledge, this galaxy has not been characterized until now. This text focuses on the study of this galaxy and provides data about it for the first time.

The galaxy WISEA J185309.64+330538.7 is situated in a relatively empty region of the sky, allowing us to estimate its distance based on its angular size without being significantly influenced by other nearby objects. At the same time, there is no known galaxy cluster in the vicinity, so in this case, as with IC 1296, we can refer to them as isolated galaxies. Simultaneously, it is possible to estimate the potential redshift of WISEA J185309.64+330538.7 from its spectrum and provide a preliminary characterization. Obtaining the spectrum of this object was particularly challenging due to its faintness, but it has ultimately been achieved with a reasonable signal-to-noise ratio.

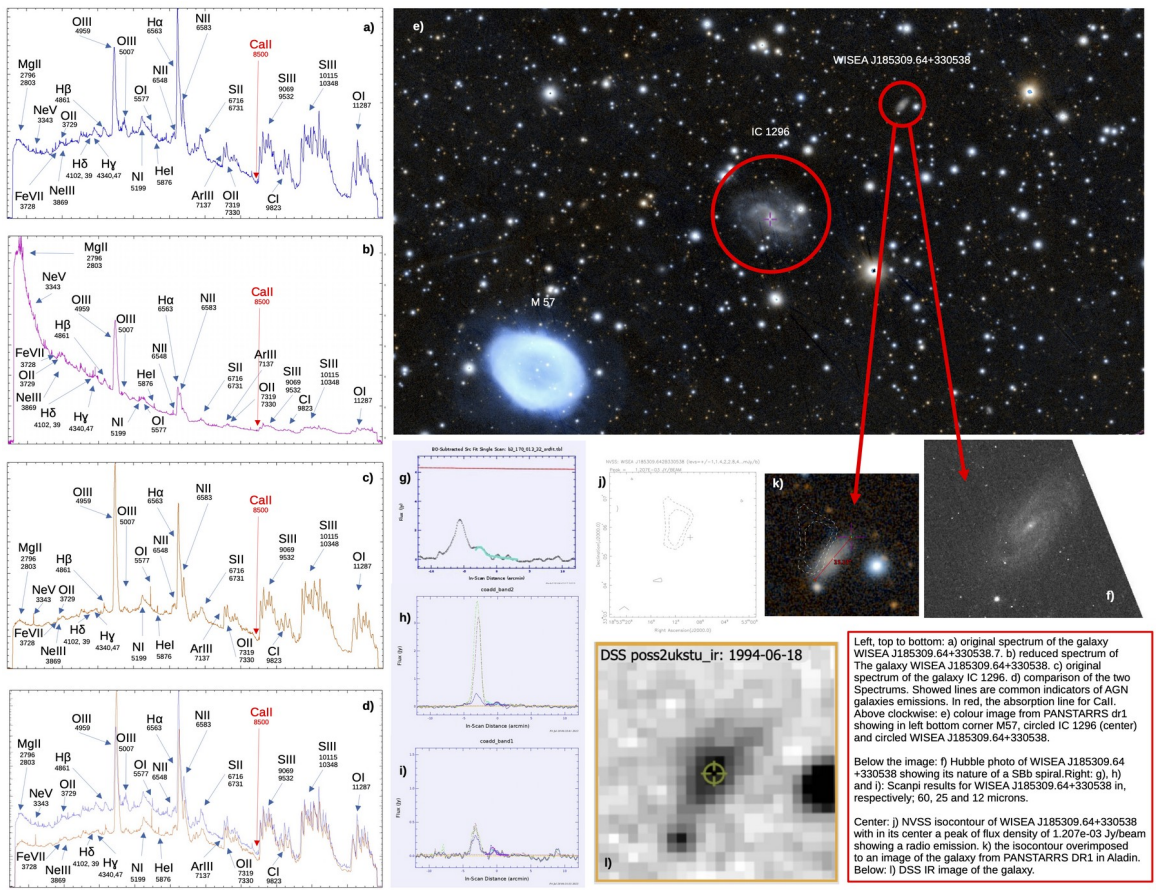

*Fig. 1. The spectra for both galaxies with emission lines noted (a, b, c, d), the two galaxies marked in an image from PANSTARRS dr1 (e), WISEA J185309.64+330538.7 in the HST image (f), the Scanpi results for 12, 25, and 60 micrometers (g, h, i) for IC 1296, the radio emission of WISEA J185309.64+330538.7 (j, k), and the DSS image for WISEA J185309.64+330538.7 (l).*

---

3   SSTSL2 stands for Spitzer Space Telescope Source List Version 2.
4   Lyon-Meudon Extragalactic Database.

In the case of IC 1296, there is a flux from CHANDRA at 18:53:18.91 +33:03:59.00, measuring approximately 35,407 eV/cm2/s (X185318.88+330359.0). Additionally, there is a modest increase in infrared (IR) emissions at 12, 25, and 60 microns. In some instances, the 12-micron emission may be linked to the presence of an Active Galactic Nucleus (AGN), while the 25 and 60 microns emissions could signify the presence of warmer and cooler dust, respectively, attributed to AGN activity. These emissions are also consistent with other possibilities, such as star formation. According to Wien's Law, dust with higher temperatures emits more strongly at shorter wavelengths. The substantial emission at 25 microns suggests relatively warm dust, while significant emission at 60 microns indicates relatively cool dust.

The spectrum of WISEA J185309.64+330538.7, obtained alongside that of IC 1296 on July 25, 2023, at the Greek Skinakas Observatory, was reduced from flats and bias frames by the author using IRAF[5] software. However, it was also desired to compare the spectrum before reduction due to the information it contains in terms of line width and height, which are important for its characterization.

An average value for nH in the vicinity of WISEA J185309.64+330538.7 reveals a high value of nH in the central region of the spiral galaxy (7.21E+20), indicating a particularly high density of hydrogen gas in this area. This could be attributed to factors such as ongoing star formation or the presence of a galactic nucleus with an active supermassive black hole, among others. The spectral lines show indications of it being a Seyfert type 1, as can be observed in Figures 1a and 1b.

The author obtains these averaged differences in the spectrum lines of WISEA J185309.64+330538.7:

- Arc: 1050 corresponds to spectrum 1140Å corresponding to 6548Å, corresponding to z=0.0857.
- Arc: 1375 corresponds to spectrum 1460Å corresponding to 4559Å, corresponding to z=0.0618.

We obtain ($H_o$ = 69.6, $\Omega_M$ = 0.286, $\Omega_{vac}$ = 0.714) an average of z=0.0737± 0.0221 (a z=0.1077 obtained through angular size falls within the error, giving us an upper limit) with a distance of approximately 291 ± 89 Mpc and a recession velocity of 22,062.49 km/s. The galaxy would have an approximate size of 16 ± 5 kiloparsecs (with an angular size of 0.00321 to 0.0042 degrees; 11.57 to 15.13 arc seconds, as seen in Figure 1k from Aladin). The orientation of the IC 1296 galaxy in Vizier[6], calculated from the angles of the J, H, K filters at 5σ, gives us 48.941 degrees, which is consistent with an orientation relative to us between 45 and 75 degrees, characteristic of Seyfert type 1.5 galaxies. Using the orientations predicted by AGN models (Bianchi et al., 2012), the orientation of WISEA J185309.64+330538.7 with respect to us should be equal to or greater than 75 degrees if it were a Seyfert type 1 galaxy.

In the case of the galaxy IC 1296, its spectrum displays (Figures 1c and 1d) subtle indications related to a Seyfert type 1.5 galaxy (Hβ and Hγ weaker than in Seyfert 1, other broad lines are weaker or more blended with the narrow lines; in addition, the Ca II 8500 Å absorption line is more likely to be deeper or more prominent in Seyfert 1.5 galaxies compared to Seyfert 1 galaxies; also, MgII is a fainter line more noticeable in 1.5 galaxies).

The unified model of AGNs indicates that the relative position of a galaxy with an active nucleus to the observer shows differences in their spectra, ranging from Seyfert-type galaxies to quasars or radio galaxies. Therefore, the position of IC 1296 may be partly responsible for receiving some spectral lines with higher intensity. Seyfert 1.5 galaxies are considered a bridge between types 1 and 2. In these galaxies, the dust torus formed around the central black hole may not be as dense or may be less aligned with our line of sight, allowing us to see both the central nucleus and the hot gas to varying degrees. This results in a spectrum with narrower emission lines than Seyfert type 1, but broader than Seyfert type 2.

---

5    IRAF is distributed by the National Optical Observatories, operated by the Association of Universities for Research in Astronomy, Inc., under cooperative agreement with the National Science Foundation.

6    https://vizier.cds.unistra.fr/viz-bin/VizieR-5?-ref=VIZ64fc954856d15&-out.add=.&-source=VII/233/xsc&recno=1233879

Regarding WISEA J185309.64+330538.7, here are 20 emission lines that can be found in the spectrum of a Seyfert type 1 galaxy (Osterbrock, 1981; Osterbrock & Shuder, 1982; Osterbrock & Dahari, 1983), with an (*) indicating the lines displayed in the spectra (Figures 1a and 1b):

1. H-alpha (Hydrogen-alpha, Hα) (*)
2. H-beta (Hydrogen-beta, Hβ) (*)
3. H-gamma (Hydrogen-gamma, Hγ) (*)
4. H-delta (Hydrogen-delta, [Hδ] λ4102) (*)
5. Lyman-alpha (Lyα)  (UV, out of reach of the spectra)
6. Lyman-beta (Lyβ)  (UV, out of reach of the spectra)
7. O III (Oxygen III, [O III] λ5007) (*)
8. O III (Oxygen III, [O III] λ4959) (*)
9. N II (Nitrogen II, [N II] λ6584) (*)
10. S II (Sulfur II, [S II] λ6717 and λ6731) (*)
11. Ne III (Neon III, [Ne III] λ3869) (*)
12. Mg II (Magnesium II, Mg II λ2798) (*)
13. Fe II (Iron II, multiple lines)  (out of reach of the spectra)
14. C IV (Carbon IV, C IV λ1549) (out of reach of the spectra)
15. C III] (Carbon III], [C III] λ1909) (out of reach of the spectra)
16. Si IV (Silicon IV, Si IV λ1397) (out of reach of the spectra)
17. He I (Helium I, multiple lines) (*)
18. He II (Helium II, He II λ4686) (*)
19. Ar III (Argon III, [Ar III] λ7136) (*)
20. Ca II (línea de absorción en λ8500) (*)
21. FeVII (Iron VIII, [FeVII] λ3728) (*)
22. OII (Oxygen II [OII], λ3729) (*)

These emission lines result from different atomic transitions and provide insights into the ionization and excitation conditions in the gas surrounding the supermassive black hole at the nucleus of the Seyfert galaxy. The author has successfully identified 16 out of the 22 lines listed, with the remaining 6 lines falling below the spectral width of the two galaxies.

With this data, the galaxy WISEA J185309.64+330538.7 could be an AGN of Seyfert 1 type, and IC 1296 an AGN of Seyfert 1.5 type, all subject to confirmation or refutation by future observations.

Furthermore, a radio emission zone is also observed in WISEA J185309.64+330538.7, as can be seen in the graph in Figure 1j. In Figure 1k, we can observe this emission zone superimposed on the galaxy image, in its corresponding spatial coordinates. A Seyfert galaxy can also be a radio galaxy by emitting radiation in the radio range due to activity in the galactic nucleus. The particle jets generating radio emission in a radio galaxy can result from processes related to the supermassive black hole at its center. The structure, reflected in Figure 1j, shows an area with a peak flux density of 1.207e-03 Jy/beam, offset from the center of the galaxy, which is consistent with the position of the spiral relative to us (more than 75 degrees of inclination within the model for Seyfert 1-type galaxies, as mentioned earlier). Figure 1l displays an infrared image from the POSS2/UKSTU1 IR[7] Database obtained at Irsa[8], revealing significant emission in the galactic nucleus, likely originating, among other areas, from the accretion disk of the potential supermassive black hole located at the galactic center.

If it were an active galaxy, this structure, likely larger than the galaxy itself, would be consistent with the remnant of a jet emitted from one of the poles of the supermassive black hole at the center of the galaxy. One would expect a symmetrical jet emitted from the other pole, which would not be visible due to the orientation of the galaxy relative to us, as the galaxy's own spiral structure of galactic matter obscures it, preventing its visibility in radio. It may also not be visible for other reasons, such as the structure of the jet itself and how it interacts with its surroundings in the intergalactic medium.

CONCLUSION

The author, after studying the spectra of the two presented objects, concludes that the previously uncharacterized galaxy WISEA J185309.64+330538.7, due to its low brightness, is of SBc barred spiral type and may be an active galaxy with an AGN, specifically a Seyfert type 1, with a redshift of $z=0.0737 \pm 0.0221$ (error obtained from the calculated redshift from the spectrum) at a distance of $291 \pm 89$ Mpc and a recession velocity of 22,062.49 km/s. The galaxy would have an approximate size of $16 \pm 5$ kiloparsecs, with its own emission characteristic of a radio galaxy, which is consistent with an active galaxy, possibly remnants of a relativistic jet typically associated with certain AGN galaxies. Furthermore, the author confirms the AGN status of the IC 1296 galaxy, most likely being a Seyfert type 1.5 galaxy.

ACKNOWLEDGEMENTS

The author would like to express gratitude to Dr. Pablo Reig, the Scientific Operations Manager of Skinakas Observatory, and Mr. Vangelis Pantoulas, an operator at Skinakas Observatory, for their invaluable contributions in obtaining and analyzing the spectra presented in this work. Furthermore, heartfelt thanks are extended to the Universidad Internacional de Valencia (VIU), especially to Dr. Elisa Nespoli, Dr. Pedro Viana Almeida, and Dr. Vicent Martínez Badenes, for their exceptional support, organization, and management of these observations as part of the M.Sc. in Astronomy and Astrophysics program.

This research has utilized the Aladin Sky Atlas developed at CDS, Strasbourg Observatory, France; the NASA/IPAC Extragalactic Database (NED); the VizieR catalogue access tool; the Pan-STARRS1 Surveys (PS1); the Two Micron All Sky Survey (2MASS); the SDSS catalogue; the Simbad astronomical database; the STScI Digitized Sky Survey; the European Space Agency (ESA) Gaia mission; and the Heasarc database. This research is based on observations obtained with the NASA/ESA Hubble Space Telescope from the Space Telescope Science Institute.

This research has utilized the NASA/IPAC Infrared Science Archive, funded by the National Aeronautics and Space Administration and operated by the California Institute of Technology.

---

7  https://archive.stsci.edu/cgi-bin/dss_form
8  https://irsa.ipac.caltech.edu/applications/finderchart/?__action=table.search&options=%7B%22tbl_group%22%3A%22upload-table-id%22%2C%22pageSize%22%3A100%7D&request=%7B%22selectImage%22%3A%22dss%2Csdss%2C2mass%2Cwise%2Ciras%22%2C%22tbl_id%22%3A%22upload-table-id%22%2C%22searchCatalog%22%3A%22yes%22%2C%22UserTargetWorldPt%22%3A%22283.2900417%3B33.09408333%3BEQ_J2000%22%2C%22id%22%3A%22QueryFinderChartWeb%22%2C%22imageSizeAndUnit%22%3A%220.08333333333333333%22%2C%22META_INFO%22%3A%7B%22tbl_id%22%3A%22upload-table-id%22%7D%2C%22pageSize%22%3A100%2C%22startIdx%22%3A0%7D